\documentclass[10pt,letterpaper]{article}
\usepackage[top=0.85in,left=2.75in,footskip=0.75in]{geometry}

\usepackage{amsmath,amssymb}

\usepackage{changepage}

\usepackage{textcomp,marvosym}

\usepackage{cite}

\usepackage{nameref,hyperref}

\usepackage[right]{lineno}

\usepackage[nopatch=eqnum]{microtype}
\DisableLigatures[f]{encoding = *, family = * }

\usepackage[table]{xcolor}

\usepackage{array}

\newcolumntype{+}{!{\vrule width 2pt}}

\newlength\savedwidth

\raggedright
\usepackage[aboveskip=1pt,labelfont=bf,labelsep=period,justification=raggedright,singlelinecheck=off]{caption}

\makeatletter
\renewcommand{\@biblabel}[1]{\quad#1.}
\makeatother

\usepackage{lastpage,fancyhdr,graphicx}
\usepackage{epstopdf}
\fancyheadoffset[L]{2.25in}
\fancyfootoffset[L]{2.25in}
\begin{document}
\vspace*{0.2in}

\begin{flushleft}
{\Large
\textbf\newline{The Structural Influence of Low-Credibility Narratives During the COVID-19 Vaccine Rollout} 
}
\newline
\\
Lynnette Hui Xian Ng\textsuperscript{1a*},
Wenqi Zhou\textsuperscript{1b},
Kathleen M. Carley\textsuperscript{2a},
\\
\bigskip
\textbf{a} Carnegie Mellon University, Pittsburgh, Pennslyvania, USA
\\
\textbf{b} Duquense University, Pittsburgh, Pennslyvania, USA
\\
\bigskip

%
%

\textsuperscript{1}These authors share first authorship.
\\
*lynnetteng@cmu.edu

\end{flushleft}
\section*{Abstract}
This work examines the structural influence of low-credibility narratives and the comparative role of automated accounts (bots) versus human users on social media platforms. To more accurately quantify the structural influence of a narrative on social media, this study proposes two novel metrics: (1) Appeal, which measures the network-weighted popularity of a message; and (2) Scope, which measures an author's message popularity-weighted network penetration.
Applying these metrics, this study analyzes 5.8 million messages from X that contain low-credibility narratives regarding COVID-19 vaccine across three distinct temporal stages: Pre-Vaccine, Vaccine Launch, and Post-Launch. 
The results demonstrate that across all timeframes, human-distributed low-credibility narratives achieved higher structural influence compared to those generated by automated accounts. Furthermore, statistical analysis reveals a significant conditional temporal effect: human-driven low-credibility narratives attained their highest Appeal and Scope during the focal Vaccine Launch week, whereas automated accounts maximized their Appeal and Scope during the highly uncertain Pre-Vaccine period. These findings highlight the distinct operational capacities of automated and organic accounts, illustrating how the Appeal and Scope of low-credibility narratives is moderated by the lifecycle stages of critical public events.


\section{Introduction}
In recent years, the dissemination of low-credibility narratives has surged during major public events, particularly during global public health crises such as the COVID-19 pandemic \cite{pierri2023one,unlu2025unveiling}. Low-credibility narratives refer to claims that contradict established scientific consensus or lack verifiable empirical support \cite{ecker2022psychological}. The uncertainty surrounding these unprecedented large-scale events creates an informational environment highly susceptible to the rapid diffusion of unverified claims or controversial narratives \cite{zhao2024understanding}. The spread of low-credibility information is a historically established phenomenon spanning traditional media and interpersonal communication \cite{broda2024misinformation}. On social media, a proliferation of automated accounts (bots) increasingly influences the spread of information \cite{ferrara2016rise,gorwa2020unpacking}.

Operating at a structurally lower cost and with substantially greater frequency than organic human users, automated bot accounts are programmed to sustain high volumes of activity \cite{schuchard2019bot,khaund2021social,ng2025global}. These programmatic entities can rapidly disseminate low-credibility content by leveraging platform distribution mechanisms to increase visibility, thereby projecting an amplified sense of public consensus \cite{yang2024anatomy,tomassi2024mapping}. Consequently, understanding the contemporary information ecosystem requires systematically distinguishing the dissemination patterns of automated systems from the organic communication behaviors of human users.

This study advances the empirical measurement of online communication and the comparative analysis of automated versus human accounts through two core contributions. In particular, we address three research questions. First, \textbf{(RQ1) Can the structural influence of narratives be methodologically identified?} Prior research frequently relies on aggregate engagement statistics, such as raw retweets, likes, and reply volumes, to quantify the narrative’s impact \cite{deverna2024identifying,van2024less,suh2010want}. While these metrics are useful indications of audience interactions with online posts, they primarily capture localized attention and often conflate dense echoing behaviors within homophilous clusters with true expansive network reach \cite{gonzalez2011dynamics,bakshy2015exposure,goel2016structural}. To provide a more robust assessment of structural influence, we integrate both message-level and network-level attributes to construct two novel, structurally weighted metrics: Appeal and Scope. Appeal captures message engagement by the relative structural prominence of the message's author in the network, while Scope weights the author's network reach by the relative popularity of the message. Unlike conventional social listening reach metrics, which typically measure the total number of unique users exposed to a brand's content, Scope weights the network position by the message momentum, capturing a narrative's potential to propagate beyond its immediate audience. Together, these two metrics capture the breadth and depth of  narrative diffusion of a message. In this study, we utilized communication data related to the COVID-19 vaccine to demonstrate the use of these composite metrics.

Second, this research empirically investigates the comparative efficacy of automated accounts and human users in distributing low-credibility information. Specifically, \textbf{(RQ2) How do bot and human users compare in the structural influence of their messages?} Further, because crisis informatics literature establishes that information diffusion is highly sensitive to the temporal progression of an event \cite{farr2025decomposing,norris2022people,lai2020pandemic}, we structured our analysis across three distinct time periods with respect to the COVID-19 vaccine release in 2020-2021  (Pre-Vaccine, Vaccine Launch, and Post-Vaccine), and ask: \textbf{(RQ3) How do the dissemination capacities differ across distinct temporal stages of a critical public event?}

We conducted descriptive analyses and estimated temporal effects using a Tweedie regression model \cite{mccullagh2019generalized} on over 5.8 million messages containing low-credibility narratives on X. The descriptive results indicate that the proportion of bot accounts participating in the discourse increased as the event progressed. 
We find that low-credibility narratives have more Appeal than non-low-credibility narratives, even when posted by the same type of user. This suggests that in this context, the audience is disproportionately receptive to low-credibility narratives, and is often willing to engage with them irrespective of whether the originating account is automated or organic.

Regression results reveal that the structural influence of these account identities is highly contingent upon the specific temporal period of the public event. We also observed divergent trajectories: automated  accounts achieved their highest levels of Appeal and Scope during the Pre-Vaccine period, a phase characterized by heightened public ambiguity and information scarcity. In contrast, organic human accounts maximized the Appeal and Scope of their narratives during the Vaccine Launch week, aligning with the release of authoritative updates and the onset of collective sense-making. By demonstrating that automated and organic actors exploit different phases of a crisis, these findings highlight the necessity for temporally dynamic platform moderation. Effective mitigation requires anticipating bot-driven amplification during early ambiguity and shifting to address human-driven dissemination during critical event milestones. 

\section{Background}
\subsection{Low-Credibility Narratives During Public Health Crises}
The emergence of global public health crises routinely exhibits a surge in the volume and influence of low-credibility information. The ambiguity and heightened public anxiety surrounding unprecedented crises promote the rapid spread of unverified information \cite{difonzo2007rumor,ng2020analyzing}. Such information has historically propagated through traditional media channels and word-of-mouth interactions \cite{katz2017personal,posetti2018short}. The propagation dynamic is a collective sense-making process where the circulation of unverified, low-credibility claims is proportional to the topic's importance and the public's uncertainty regarding the facts \cite{allport1947psychology}.

However, the architecture of online platforms has fundamentally changed the velocity, scale, and mechanisms by which these narratives diffuse digitally. Over the past decade, the COVID-19 pandemic has served as the most salient paradigm of this shift. The unprecedented nature of the virus and the initially fragmented scientific consensus created significant informational voids, which were rapidly filled by alternative explanations and contested health directives \cite{purnat2021infodemic}. Survey data from the Pew Research Center in 2020 indicated that approximately 48\% of Americans reported encountering COVID-19-related low-credibility information online\cite{jurkowitz2020early}. The ``infodemic" characterized by  the rapid proliferation of low-credibility narratives during the pandemic was also fueled by automated accounts that spread narratives to massive audiences before authoritative and scientific consensus could be established \cite{shao2018spread}.

\subsection{Measuring Influence of Online Information}

A substantial body of literature evaluating the diffusion of low-credibility information relies primarily on aggregate engagement statistics, such as retweet counts, likes counts, and reply volumes \cite{van2024less,suh2010want}. These metrics have been widely adopted across disciplines because they are readily accessible, highly quantifiable, and provide immediate, valuable signals of user attention and content virality \cite{cha2010measuring,kwak2010twitter,goel2016structural}. These metrics are also intuitive. For example, a higher retweet count could indicate that a tweet has more influence across the network \cite{deverna2024identifying}. By capturing the direct interactions between users and content, traditional engagement metrics offer a baseline for assessing the immediate popularity of a message within a platform's ecosystem. 

However, while these metrics are excellent indicators of localized engagement, they do not capture the structural dimension of influence dissemination. High engagement volume reflects immediate attention but may not fully delineate the structural dissemination of a message across the broader network. A message may accumulate a high number of retweets within a dense, homophilous cluster of users, thereby exhibiting significant localized popularity within the author's ego-network, but might have constrained network-wide reach beyond the ego's alters \cite{gonzalez2011dynamics,bakshy2011everyone}. Consequently, while message-level attributes of engagement are methodologically sound measures, they capture only direct user interaction and not the expansive structural influence of a message.

To capture a more comprehensive measure of influence, scholarship has increasingly emphasized the value of examining the network-level attributes of content creators alongside message characteristics. Research indicates that structural positions, such as an author's degree centrality, play a pivotal role in determining the potential reach of narratives \cite{aral2012identifying,borge2012locating}. Users embedded within highly interactive sub-networks possess the capacity to disseminate content extensively through networks built over time, even if their baseline engagement metrics appear modest \cite{lee2023finding}. Instead of treating message popularity and the author's network position as purely disjointed metrics, we view them as interacting dimensions of influence, and construct multi-dimensional metrics that link message popularity with network structure.

\subsection{Bots vs Humans in Online Events}
To accurately map the diffusion of low-credibility narratives, it is necessary to differentiate the sources of online communication. Specifically, current literature distinguishes between human users and automated bot accounts. Bot accounts are software-driven entities programmed to generate content and interact within social media networks, and are often identified through behavioral analysis with machine learning classifiers \cite{zhang2025research,ng2023botbuster}. This operational distinction is salient because the two account types function under different structural and volume constraints. Human communication is inherently bounded by physical time, attention capacity, and the limits of pre-existing social ties \cite{pennycook2022nudging,miritello2013limited}. In contrast, automated bot accounts can operate without equivalent constraints, allowing them to sustain high-frequency posting schedules and interact at scale \cite{bessi2016social,ng2025global}. By maintaining elevated activity levels, automated accounts can increase the baseline visibility of specific messages within platform distribution systems, thus accelerating the volume-based dissemination of low-credibility information \cite{broniatowski2018weaponized,shao2018spread}. Consequently, measuring online content engagement requires separating organic human communication networks from high-volume, automated distribution patterns.

Beyond account type, the distribution of information is moderated by temporal factors, particularly during critical public events \cite{arif2016information,lai2020pandemic}. Crisis informatics literature establishes that public events, such as public health crises, progress through distinct temporal phases characterized by varying levels of data availability and public inquiry \cite{palen2017social}. The informational environment shifts considerably as an event moves from early anticipation to the occurrence of focal events, and eventually into resolution or adoption phases \cite{starbird2014rumors,compton2020temporality,samimian2022scenarios}. For instance, the discourse surrounding the COVID-19 vaccine rollout transitioned through a pre-vaccine period of high anticipation, the immediate focal event of the vaccine launch, and a post-vaccine period of early adoption \cite{blane2022social}.

Empirical observations indicate that account activity levels demonstrate measurable variation across these distinct stages. During initial phases, where authoritative information is often sparse and uncertainty is high, automated accounts tend to exhibit increased activity, distributing low-credibility narratives prior to the establishment of official consensus \cite{himelein2021bots,pierri2023one}. In contrast, organic human engagement typically peaks during focal events, such as official policy announcements or verifiable product releases, when empirical developments stimulate increased user discussion and information sharing \cite{ng2020analyzing}.

While the distinct operational capacities of human and automated accounts are documented, alongside their temporal variations in activity, limited empirical work quantifies how the structural influence of these distinct account types compares across the lifecycle of a specific public event. This research addresses this gap by assessing the interaction between account identity (bot versus human) and event stage. By applying the proposed Appeal and Scope metrics across the Pre-Vaccine, Vaccine Launch, and Post-Vaccine periods, this study provides a precise evaluation of when, and to what extent, automated and human accounts maximize their network influence in the distribution of low-credibility information.

\section{Structural Influence Metrics}
To capture the multidimensional nature of information diffusion, we designed two composite metrics of structural influence that move beyond raw engagement volumes. Traditional aggregate statistics frequently conflate localized echoing with true expansive reach, because engagement counts may reflect concentrated activity within tightly connected communities rather than broad diffusion across the network \cite{gonzalez2011dynamics,deverna2024identifying}. 

To resolve this, we propose two topology-aware metrics that integrate message-level popularity with network-level structural position: \textbf{Appeal} and \textbf{Scope}. Appeal focuses on the message: its popularity weighted by the author's relative connectivity. Scope focuses on the network center: the structural reach of the author weighted by  message popularity. These metrics integrate network science research that demonstrates how the structural position of users within communication networks influences information spread \cite{barabasi1999emergence,goel2016structural}.

To calculate these metrics, we first constructed an all-communication network graph for each \texttt{TimePeriod} $t$, representing users $U$ as nodes and interactions as links. Specifically, users who tweeted during \texttt{TimePeriod} $t$ were represented as nodes in the graph. Users $u_a$ and $u_b$ were linked together if they had an interaction through a tweet, i.e., a retweet or a mention within the tweet. An all-communication graph was used to capture the entirety of the observed interactions in the network. In such a network, both direct engagement behaviors (e.g. re-tweeting) and conversational relationships (e.g. tagging another user) contribute to the structural representation of the information environment.

\subsection{Appeal Metric}
The Appeal metric captured the popularity of a message $m$ by incorporating the engagement value a message received and the structural position of the account that generated it. A message author's position in the network can significantly affect the diffusion potential of the information that they produce, because users who are more connected can disseminate content through a larger set of interaction pathways \cite{barabasi1999emergence,goel2016structural}. This also means that messages produced by structurally well-connected authors may exert broader influence even if raw engagement levels might appear low. As specified in the below \autoref{eq:appeal}, the Appeal metric weights the engagement of a message by the relative network connectedness of the author.

\begin{equation}
    \texttt{Appeal}_{m,u} = \texttt{RetweetCount}_m \times (1 + \texttt{TotalDegreePercentile}_{u,t})
\label{eq:appeal}
\end{equation}
where $\texttt{RetweetCount}_m$ is the retweet volume of message $m$, which captures the immediate popularity of the message. $\texttt{TotalDegreePercentile}_{\{u,t\}}$ is the percentile ranking of author $u$ based on their total degree in the network during that period $t$. Total degree measures the sum of the number of incoming and outgoing interactions author $u$ has. This value represents the message author's connectedness within the communication network. By combining immediate popularity with structural connectivity, the Appeal metric distinguishes between localized popularity within small clusters (i.e., echo chambers) and attention generated by structurally well-connected users that can influence the broader network because of their position.





\subsection{Scope Metric}
Scope represents the potential structural reach of a message $m$, incorporating the connectivity of its author $u$ and the observed popularity of the message itself. Network science literature shows that nodes (users) with higher degree centrality are more likely to facilitate larger information diffusion cascades because they are directly connected to a larger set of neighbors who can further propagate the information \cite{barabasi1999emergence,newman2018networks}. Therefore, messages produced by authors with high structural connectivity can potentially reach a larger portion of the network. As specified in \autoref{eq:scope}, the Scope metric weights the message's structural reach, which is largely determined by the author's network position, by the message's relative popularity.
\begin{equation}
    \texttt{Scope}_{m,u} = \texttt{TotalDegreeCentrality}_{u,t} \times (1 + \texttt{RetweetCountPercentile}_{m,t})
\label{eq:scope}
\end{equation}
where $\texttt{TotalDegreeCentrality}_{\{u,t\}}$ is the message author $u$’s  total degree centrality within \texttt{TimePeriod} $t$. This reflects the number of direct connections which information from author $u$ can potentially propagate. $\texttt{RetweetCountPercentile}_{\{m,t\}}$ is the percentile ranking of message $m$ based on its retweet count during \texttt{TimePeriod} $t$. This captures the relative popularity or ``hotness" of the message $m$ among other messages in the same time period. By combining the structural reach of the author with the relative popularity of the message, the Scope metric captures the potential diffusion breadth that a message can achieve within the network. 

The specific formulation of adding 1 to the percentile rank (which inherently ranges from 0 to 1)  ensures that the base metric (either engagement volume or network centrality) is strictly preserved as a baseline multiplier of 1. It is then proportionally amplified by up to a factor of 2 based on the weighting dimension's relative ranking, providing a mathematically stable index that penalizes neither dimension excessively.

\section{Data and Methods}
To empirically evaluate the structural influence of narratives, we executed a multi-stage methodological pipeline. This involved curating a large-scale platform dataset, utilizing semantic vectorization (with TwHIN-BERT \cite{zhang2023twhin}) to classify narrative types, applying algorithmic probability scoring (with BotHunter \cite{beskow2018bot}) to differentiate bot accounts from human users, and constructing formal engagement metrics. We then estimated conditional temporal effects using a compound Poisson-Gamma distribution model to quantify how the structural influence of two distinct user account types (bot and human) varies across the different temporal stages of the vaccine rollout.

\subsection{Data Curation and Pre-processing}
We adopted a published dataset of social media discussions regarding the COVID-19 vaccine, hereafter referred to as CovidInfo \cite{blane2022social}. To facilitate temporal analysis, the CovidInfo data was divided into three distinct periods: Pre-Vaccine (December 1–7, 2020), Vaccine Launch (December 8–10, 2020), and Post-Vaccine (January 25–31, 2021). These periods capture discussions on X centered around the initial COVID-19 vaccination rollout, amounting to about 8.6 million tweets. This serves as an appropriate case study because of its clear demarcations of temporal stages and the documented prevalence of low-credibility narratives \cite{pierri2023one,blane2022social,ng2020analyzing,himelein2021bots}.

\subsection{Identifying Low-Credibility Narratives}
We first needed to isolate low-credibility narratives in our corpus. To do so, we filtered the CovidInfo corpus against an expertly annotated reference dataset \cite{memon2020characterizing}. The reference dataset contains manual categorizations of low-credibility COVID-19 narratives (i.e., fake cure, conspiracy, fake treatment, false fact or prevention, and false public health response) from the 2020 pandemic. The temporal alignment of this dataset with our CovidInfo dataset ensures narrative consistency. We first pre-processed messages by removing artifacts such as URLs and user mentions. We then generated dense vector representations of the texts using TwHIN-BERT embeddings \cite{zhang2023twhin}. The TwHIN-BERT embeddings were trained directly by X's research team using its proprietary recommendation algorithm \cite{x_twitter_recommendation_2023}; therefore, they are well-suited for capturing domain-specific semantic relationships.

To validate the embeddings, we trained a multi-class logistic classifier on a randomly selected 80\% subset of the aforementioned reference data, using the embeddings as input and the annotated narrative groups as output. The classifier achieved an accuracy of 50.13\% on the remaining 20\% holdout set, exceeding the random-chance baseline of 20\%. We then classified the unannotated messages in CovidInfo using an all-pairs cosine similarity procedure adapted from prior text-matching studies \cite{ng2022cross,ng2021coronavirus}. Specifically, all reference messages $v_1$ to $v_n$ from the reference dataset were compared with an unannotated message $w$ in our dataset. For each unannotated message $w$ and reference message $v$, we computed the cosine similarity $S_{w,v}$ between their respective embeddings. An unannotated message $w$ was assigned to the corresponding narrative group of $v$ if their cosine similarity score $S_{w,v}\geq0.7\in[0,1]$. This indicates that the unannotated message $w$ is at least 70\% semantically similar to the known message $v$. This threshold operates on the assumption that high semantic resemblance to the manually curated dataset constitutes a true positive match. The final dataset comprises 5,890,967 messages classified as low-credibility narratives from CovidInfo dataset. Anonymized examples of  matched messages and low-credibility narratives are provided in Supporting Information \autoref{supp:matching_messages} and Supplementary Materials \autoref{supp:narrative_groups}. 

\subsection{Bot Account Annotation}
To classify account types, we applied BotHunter, a tier-based automated account detection model. BotHunter consists of several hierarchical Random Forest classifiers trained on manually annotated social media data \cite{beskow2018bot}. The algorithm evaluates account features, such as username, post texts, and metadata (e.g., number of followers and number of likes), to calculate an automation probability score,  $P(bot)$, ranging from 0 to 1. BotHunter has demonstrated classification accuracy exceeding 90\% in standard benchmark evaluations \cite{beskow2018bot}. It has been extensively validated across various sociotechnical contexts, including geopolitical discourse \cite{alieva2022investigating}, health pandemics \cite{phillips2025emotions}, and civic elections \cite{jacobs2022whodefinesdemocracy}.

To prioritize precision, accounts with $P(bot)\geq0.70$ were classified as bots, while those with $P(bot)<0.70$ were classified as humans. This specific threshold relies on a prior large-scale  statistical analysis demonstrating the longitudinal stability of these probability scores \cite{ng2022stabilizing}. Applying this configuration to our dataset, 519,337 accounts (23.1\%) were classified as automated, producing 41.2\% of the total messages. The remaining 1,738,645 accounts were classified as human users.

\subsection{Empirical Evaluation}
To empirically evaluate the structural influence of low-credibility narratives distributed by distinct account types across varying temporal stages, we constructed two regression models for our two structural influence metrics.
The key dependent variables, $\texttt{Appeal}_m$ and $\texttt{Scope}_m$, are represented as $\texttt{Metric}_m$ in the general equations that were applied to both metrics.  As indicated in the subsequent section, both exhibit a high frequency of exact zeros and a heavily right-skewed continuous tail. Both retweet counts and network degree centrality follow heavy-tailed distributions in social media networks, where a small number of users or messages receive disproportionately large levels of attention and connectivity \cite{barabasi1999emergence,goel2016structural}. Therefore, we estimated these models via a Tweedie regression method with a variance power parameter of $p=1.5$, which corresponds to a compound Poisson-Gamma distribution. This model accounts for the non-negative, zero-inflated and heavily right-skewed distributions of the Appeal and Scope metrics. 

\paragraph{Baseline Model}
The first specification is a baseline model designed to isolate the independent main effects of account identity and temporal period, while controlling for message-level and user-level covariates:

\begin{equation}
\begin{aligned}
\texttt{Metric}_m =\;&
\alpha_0 
+ \alpha_1 \texttt{IsBot}_u
+ \alpha_2 \texttt{TimePeriod}_m
+ \alpha_3 \texttt{NarrativeGroup}_m \\
&+ \alpha_4 \texttt{IsRetweet}_m
+ \alpha_5 \texttt{AccountAge}_u
+ \varepsilon_i
\end{aligned}
\label{eq:baseline_model}
\end{equation}

\paragraph{Conditional Effect Model}
The second specification introduces an interaction term to capture the conditional temporal effects, testing whether the engagement of automated accounts vs human accounts fluctuates depending on the lifecycle stage of the public event:

\begin{equation}
\begin{aligned}
\text{Metric}_m =\;&
\beta_0
+ \beta_1 \texttt{IsBot}_u
+ \beta_2 \texttt{TimePeriod}_m
+ \beta_3 (\texttt{IsBot}_u \times \texttt{TimePeriod}_m) \\
&+ \beta_4 \texttt{NarrativeGroup}_m
+ \beta_5 \texttt{IsRetweet}_m
+ \beta_6 \texttt{AccountAge}_u
+ \delta_i
\end{aligned}
\label{eq:conditional_model}
\end{equation}

In both models, $m$ denotes an individual message and $u$ denotes its account owner. The primary independent variables are $\texttt{IsBot}_u$, a binary indicator of whether the user's bot probability score exceeded the 0.70 threshold, and $\texttt{TimePeriod}_m$, a categorical variable distinguishing the Pre-Vaccine, Vaccine Launch, and Post-Vaccine phases.

For robust estimates, the model incorporates several control variables. $\texttt{NarrativeGroup}_m$ represents the specific thematic categorization of the low-credibility claim (e.g., fake cure, conspiracy), as semantic framing inherently affects engagement. $\texttt{IsRetweet}_m$ controls for whether the message was an original tweet or a retweet, which influences the recommendation from the platform algorithm. Finally, $\texttt{AccountAge}_u$ was calculated as the duration between the creation date of tweet $m$’s account $u$ and the last date of the $\texttt{TimePeriod}_m$. It controls for the cumulative accumulation of user engagement over time. Variance inflation factor (VIF) scores were calculated for all predictors for each model, returning values strictly below the conventional threshold of 5 \cite{kutner1984applied}, indicating no problematic multicollinearity. The full results of the VIF values are presented in the Supplementary Information \autoref{supp:vif_values}.

\section{Results}
\subsection{Descriptive Analysis}
Analysis of the original CovidInfo dataset indicates that bots constitute an average of $26.70\%(\pm0.82\%$) of the active user population per period. This prevalence is consistent with prior empirical studies examining automated accounts and their behaviors within coronavirus vaccination discussions on X, which generally report baseline bot proportions between 15 and 20\% \cite{ng2025global,zhang2023social,suarez2022assessing}.
Supplementary Information \autoref{tab:distribution_account_identities} presents the detailed data statistics of each user type across all time periods.

The empirical evaluation of our three-period dataset reveals shifting distributions of low-credibility narratives across the lifecycle of the public health event. Descriptive analyses indicate that low-credibility information constituted a larger proportion of total message volume during the initial phases, specifically accounting for 69.9\% of messages in the Pre-Vaccine period and 70.7\% during the Vaccine Launch, before declining to 63.4\% in the Post-Vaccine period. At the account level, organic human participation peaked during the focal event of the Vaccine Launch, subsequently decreasing as the rollout progressed. Conversely, bots demonstrated a stable and increasing structural presence over time, resulting in an elevated ratio of automated to human accounts, from 35.0\% to 38.2\%, by the Post-Vaccine phase. Notably, across all observed periods, bots consistently produced a higher relative proportion of low-credibility narratives compared to human users. 

\begin{figure}
    \centering
    \includegraphics[width=\linewidth]{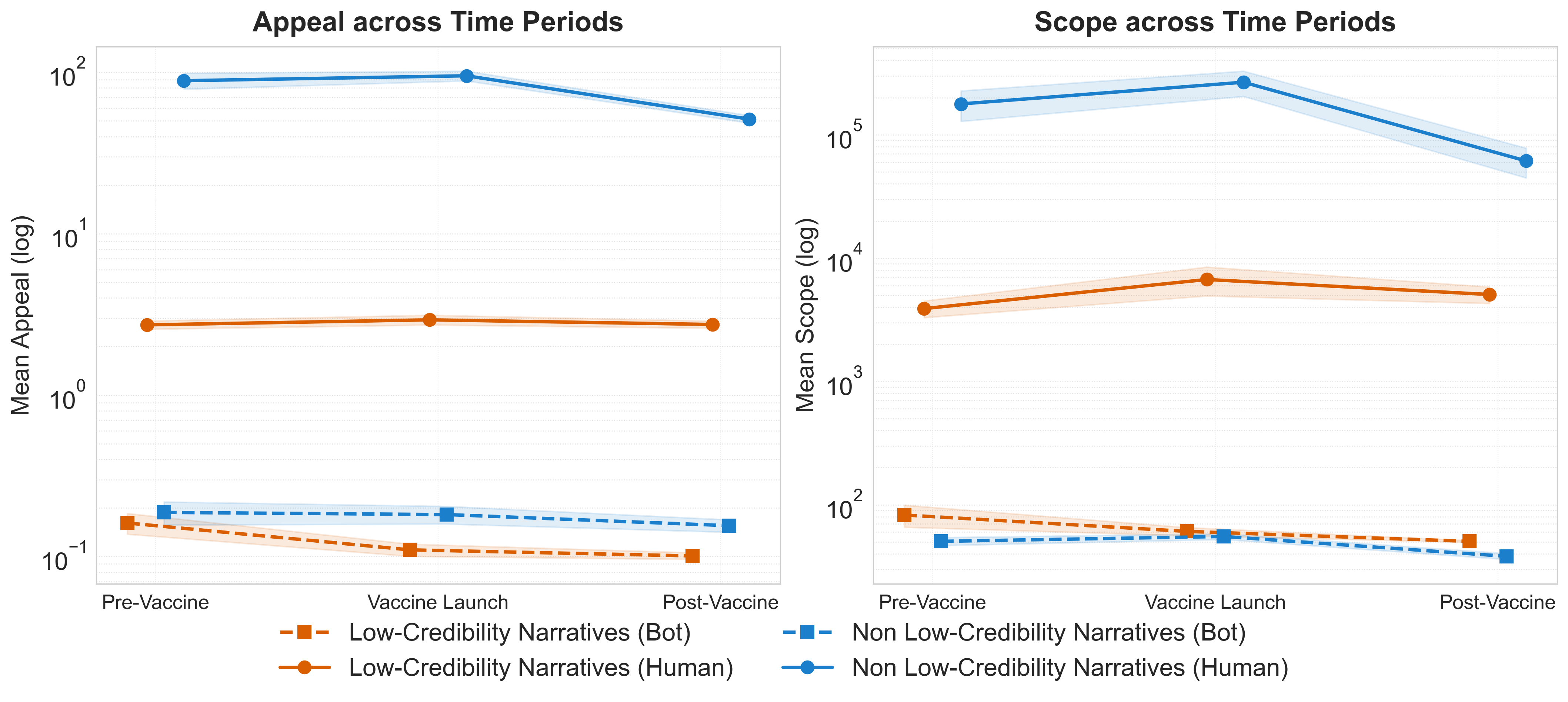}
    \caption{Average Appeal and Scope of Tweets across the three Time Periods.}
    \label{fig:results_time_period}
\end{figure}

Addressing RQ2, the difference in structural influence of bot and human users, when assessing message engagement, \autoref{fig:results_time_period} reveals that low-credibility narratives generally exhibited lower average Appeal and Scope compared to standard information distributed by the same account type (mean log(Appeal) low-credibility vs non-low-credibility narratives = 0.12 vs 0.17 for bots, 2.81 vs 76.43 for humans; log(Scope) low-credibility vs non-low-credibility narratives = 63.25 vs 47.29 for bots, 5429.98 vs 164296.51 for humans). Supplementary Information \autoref{tab:appeal_scope_desc} presents the detailed data statistics for each metric. However, a significant deviation emerged within automated networks: low-credibility narratives distributed by automated accounts were on average 34\% more widespread, measured by scope, than non-low-credibility narratives disseminated by these same automated accounts. 

Further, human users overall achieved substantially higher structural influence than automated accounts measured by both Appeal and Scope (Mann-Whitney U; Appeal: $|r| = 0.100$, Scope: $|r| = 0.071$, both $p<0.001$). However, this performance gap was compressed for low-credibility narratives. Low-credibility narratives distributed by humans generated 22 times the Appeal of their automated counterparts, whereas  non-low-credibility narratives by humans generated 449 times the Appeal of their bot counterparts. This suggests that automated accounts are disproportionately optimized for the dissemination of low-credibility content relative to standard information communication. The detailed statistics are presented in Supplementary Material \autoref{tab:narratives_distribution}.

Turning to RQ3, the different dissemination capacities of temporal stages, \autoref{fig:results_time_period} also situates the observed patterns of message Appeal and Scope across the temporal lifecycle of the vaccine rollout. For human accounts, both Appeal and Scope for low-credibility narratives rose markedly during the Vaccine Launch before declining in the Post-Vaccine period, consistent with collective sensemaking dynamics documented in crisis informatics literature\cite{farr2025decomposing,palen2017social}. In contrast, bot-distributed low-credibility narratives exhibited a comparatively flatter trajectory across all three periods, with modest increases across time. The gap in Appeal and Scope between human and bot accounts was widest during the Vaccine Launch period, and narrowest during the Pre-Vaccine period.
These descriptive temporal observations motivate the formal regression analysis presented in \autoref{sec:regression_analysis}.

To formally validate the distributional separation between bot and human users, and to confirm the stability of these group-level differences across three temporal stages, we conducted two non-parametric statistical tests. First, we applied a Mann-Whitney U test to each metric across all bot vs human accounts, irrespective of narrative type, and quantified the between-group separation using the rank-biserial correlation coefficient $r$. Second, to assess whether within-group metric distributions shift substantially over time, for each of the narrative groups, we applied a Kruskal-Wallis H-test across the three time periods, computing epsilon-squared ($\varepsilon^2$) as the effect size.

\begin{figure}
    \centering
    \includegraphics[width=\linewidth]{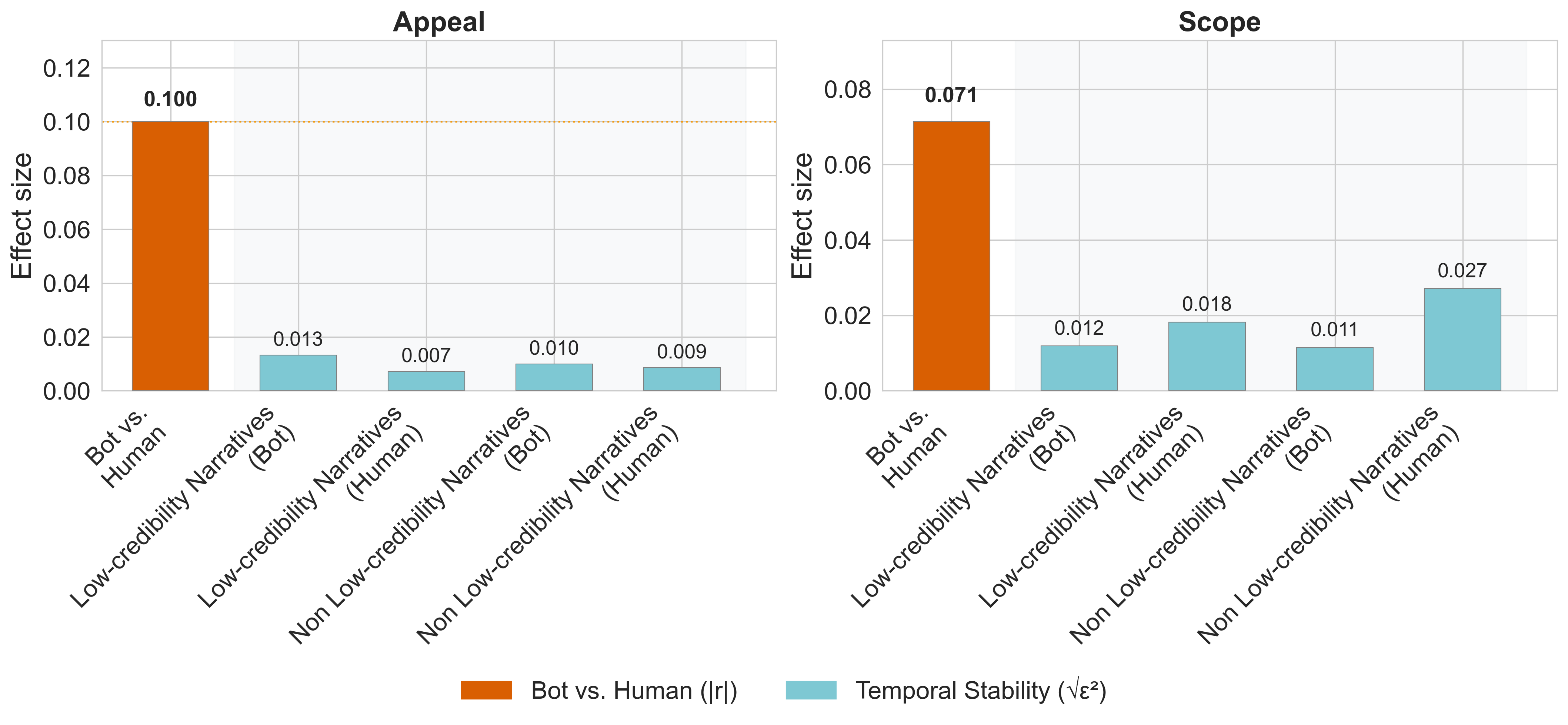}
    \caption{Effect size contrast between between-group (bot-human) separation (orange) and within-group temporal stability (blue), from Mann-Whitney U test rank-biserial coefficient $|r|$ and Kruskal-Wallis $\sqrt{\varepsilon^2}$ respectively.} 
    \label{fig:statsanalysis}
\end{figure}

\autoref{fig:statsanalysis} presents these two effect sizes as a contrast plot. For Appeal, the bot-human separation reaches $|r|=0.100$, a small-effect threshold, while temporal effects remain below  $\sqrt{\varepsilon^2} = 0.013$. For Scope, the bot-human separation ($|r| = 0.071$) similarly exceeds the temporal effects (average $\sqrt{\varepsilon^2} = 0.019$). Across both metrics, the between-group effect consistently exceeds any within-group temporal change, indicating that the structural distinction between automated and organic accounts is a stable property of their network behavior rather than an artifact of a specific temporal window. This provides empirical support for treating account type as a reliable structural predictor of Appeal and Scope in our regression model.

\subsection{Regression Analysis}
\label{sec:regression_analysis}
To formally quantify these relationships and isolate idnependent effects, a Tweedie regression model with a compound Poisson-Gamma distribution was utilized. The results of the regression are presented in \autoref{tab:regression}, with the full results in \autoref{tab:tweedie_results}. The baseline model's results indicate that the overall influence of low-credibility narratives peaked significantly during the Vaccine Launch phase. During this focal period, these narratives demonstrated a 6.92\% increase in Appeal and a 52.04\% increase in Scope compared to other periods. Furthermore, after controlling for message and network attributes, the baseline estimations confirm that human users consistently achieved broader influence through engagement and network penetration; low-credibility narratives disseminated by automated accounts exhibited 91.11\% lower Appeal and 89.56\% lower Scope compared to their human counterparts.

The conditional effect model, which introduces an interaction term between account identity and temporal period, reveals divergent behavioral trajectories (RQ3). For human users, the structural influence of low-credibility narratives was maximized during the Vaccine Launch week, characterized by a 14.11\% increase in Appeal and an 85.52\% increase in Scope. In contrast, automated accounts demonstrated an inverse temporal pattern. The interaction coefficients indicate that the influence of low-credibility narratives from automated accounts was minimized during the Vaccine Launch, reaching its maximum efficacy during the Pre-Vaccine period. While human users maintained higher overall influence across all periods than automated accounts, the performance discrepancy was most salient during the Vaccine Launch, where automated accounts exhibited 30.2\% lower Appeal and 62.3\% lower Scope relative to human baselines. These findings formally establish that the network efficacy of automated accounts is highly contingent upon the specific temporal stage of a public event.

\begin{table}[h]
\centering
\begin{tabular}{lcccc}
\hline
 & \multicolumn{2}{c}{Baseline Model} & \multicolumn{2}{c}{Conditional Effect Model} \\
\cline{2-3} \cline{4-5}
 & Appeal & Scope & Appeal & Scope \\
\hline
Bot = 1 & -2.42$^{***}$ & -2.26$^{***}$ & -2.19$^{***}$ & -1.64$^{***}$ \\
Vaccine Launch = 1 & 6.70E-2$^{***}$ & 0.42$^{***}$ & 0.13$^{***}$ & 0.62$^{***}$ \\
Post-Vaccine = 1 & 4.38E-2$^{**}$ & 0.23$^{***}$ & 9.59E-2$^{***}$ & 0.38$^{***}$ \\
Bot $\times$ Vaccine Launch &  &  & -0.36$^{***}$ & -0.98$^{***}$ \\
Bot $\times$ Post-Vaccine&  &  & -0.28$^{***}$ & -0.68$^{***}$ \\
\hline
\multicolumn{5}{l}{\footnotesize Notes: $^{***}$ p$<$0.001, $^{**}$ p$<$0.01, $^{*}$ p$<$0.05.}
\end{tabular}
\caption{Tweedie Regression Results using low-credibility Covid-19 Narratives on X. Detailed regression results are presented in \autoref{tab:tweedie_results}.}
\label{tab:regression}
\end{table}


\section{Discussion and Conclusion}
This study conceptualizes influence as a multidimensional construct that inherently links message popularity with network structure. To more accurately quantify the impact of narratives in a network, we introduce two structurally weighted metrics, \textbf{Appeal and Scope}, and apply them to evaluate the distribution of narratives during a critical global health event. Rather than relying on isolated engagement counts, these metrics integrate message-level and network-level attributes: Appeal weights a message’s engagement by its author’s structural prominence, while Scope weights an author’s network centrality by the message’s popularity. By interacting raw message engagement with topological network reach, our findings offer a new framework for understanding how human- and bot-authored narratives spread across distinct temporal stages. This framework incorporates the network structure, factoring in for the networked-constrained dependence of a message, because message appeal and scope are affected by the types of users that the author can reach \cite{ng2026social}.

A central finding of this research is that human users consistently outperformed bots in achieving network-wide influence. Although bots consistently produced a higher relative proportion of low-credibility narratives, human-distributed messages generated substantially higher Appeal and Scope. This empirical observation aligns with network science literature emphasizing the necessity of organic social capital for deep network penetration \cite{burt2005brokerage}. Bots can generate immense localized volume, but they typically lack the established, bidirectional trust networks required to bridge distinct structural clusters \cite{gonzalez2011dynamics}. Furthermore, literature indicates that human users are highly responsive to low-credibility narratives because of novelty and skepticism, often voluntarily and unwittingly accelerating their distribution through the social network\cite{vosoughi2018spread,diaz2023disinformation,ng2020analyzing}. However, while bots face structural boundaries, their operational architecture is disproportionately optimized for the dissemination of low-credibility content over standard discourse. Low-credibility narratives distributed by bots were nearly 34\% more widespread than standard information disseminated by those same accounts.

The Tweedie regression results formally establish that the network efficacy of different account types is deeply contingent upon the temporal stage of the public event. Automated accounts achieved their maximal structural influence during the Pre-Vaccine period, a phase characterized by heightened public uncertainty and a lack of official consensus. This phenomenon can be contextualized through the theoretical lens of ``data voids'' \cite{boyd2018data}. During initial anticipatory phases, authoritative information is scarce. Bots capitalize on these voids, leveraging automated high-frequency posting to establish baseline visibility for low-credibility narratives before organic human consensus can form \cite{viralityproject2024,geissler2023russian}. During the Vaccine Launch period, the structural influence of bots was substantially compressed. The negative interaction coefficients ($-30.2\%$ Appeal, $-62.3\%$ Scope relative to human baselines) indicate that the surge of authoritative information accompanying a focal event (i.e., government campaigns about vaccination benefits) could have displaced bot-driven low-credibility narratives from the structural center of the discourse. This is consistent with research showing that high-salience events generate rapid, organic information cascades that crowd out low-authority content \cite{starbird2014rumors,palen2017social}. The Post-Vaccine period reveals a partial recovery in bot structural influence. While bots remained suppressed relative to humans, the regression coefficients for the interaction terms are attenuated compared to the Vaccine Launch, suggesting that automated accounts partially reclaimed structural footing in the information environment.

Conversely, human users maximized their network influence during the focal event of the Vaccine Launch. Focal periods can be triggers for collective sensemaking and prompt widespread organic discussion, information sharing, and debate among human users \cite{palen2017social,starbird2014rumors,farr2025decomposing}. The divergence in these temporal peaks highlights a strategic transition: automated accounts establish narrative prominence during periods of ambiguity, whereas human users drive structural dissemination during periods of concrete event realization.

While our methodological framework provides a rigorous evaluation of the structural influence of online messages, several limitations must be acknowledged to contextualize the findings. First, the empirical estimations rely on specific, mathematically derived thresholds, namely the probability cutoff for automated account classification and the cosine similarity threshold for identifying low-credibility narratives. While these are grounded in prior validated literature, future research should conduct extensive sensitivity analyses across varying thresholds to assess the ultimate stability of these structural network metrics. Second, the dataset is restricted to a single microblogging platform. Because platform architecture and recommendation algorithms inherently shape communication networks, these structurally weighted metrics should be tested across alternate platforms to ensure construct validity across ecosystems. Finally, the analysis focuses exclusively on the COVID-19 vaccine rollout. While this serves as an optimal case study for a global public event, extending this methodology to non-health contexts is necessary to determine if these temporal interaction effects are universally generalizable.

Ultimately, this research demonstrates that evaluating the diffusion of low-credibility information requires methodological frameworks that account for both network topology and temporal dynamics. Relying on aggregate engagement statistics obscures the fundamentally different ways in which human users and automated accounts navigate social networks. By introducing and validating the structurally weighted metrics of Appeal and Scope, this study provides a mechanism for disentangling high-volume, automated dissemination from organic structural network penetration. The empirical divergence observed across the COVID-19 vaccine rollout lifecycle underscores a critical theoretical takeaway: automated accounts optimize for the rapid saturation of data voids during periods of high ambiguity, whereas organic human networks drive deep structural diffusion during periods of collective sensemaking. Recognizing these distinct operational boundaries and temporal dependencies is essential for both advancing network science literature and developing targeted interventions during future global public events.


\section*{Declarations}
\paragraph{AI Statement} During the preparation of this work, the author(s) used Google Gemini to refine the academic tone, improve readability (i.e., spelling and grammar), and assist with the structural organization of the manuscript (e.g. table formatting). After using this tool, the author(s) closely reviewed, thoroughly edited, and validated all content. The author(s) take full responsibility for the originality, accuracy, and integrity of the final publication.

\paragraph{Data Availability} Access to the datasets is available upon request, in accordance to the data sharing policies of X.com. 

\section*{Acknowledgments}
The first and third authors are supported by the Scalable Technologies for Social Cybersecurity, U.S. Army (W911NF20D0002), the Minerva-Multi-Level Models of Covert Online Information Campaigns (N000142112765), Threat Assessment Techniques for Digital Data (N000142412414), and MURI (N000142112749), Office of Naval Research.

%
%
%
\bibliography{references}

\clearpage
\appendix

\renewcommand{\thefigure}{S\arabic{figure}}
\renewcommand{\thetable}{S\arabic{table}}
\renewcommand{\thesection}{S\arabic{section}}

\setcounter{figure}{0}
\setcounter{table}{0}
\setcounter{section}{0}


\section{Data Statistics}
\label{supp:data_statistics}
\autoref{tab:distribution_account_identities} presents the statistics of the distribution of the user types within the data over each of the time periods. \autoref{tab:narratives_distribution} presents the distribution of low-credibility narratives within the dataset. \autoref{tab:appeal_scope_desc} presents the descriptive statistics of the metrics for the dataset.

\begin{table}[h]
\centering
\begin{tabular}{lccc}
\hline
\textbf{TimePeriod} & \textbf{Bots} & \textbf{Humans} & \textbf{Total Users} \\
\hline
Pre-Vaccine & 153,379 (26.44\%) & 426,756 (73.56\%) & 580,135 \\
Vaccine Launch& 220,991 (26.04\%) & 627,499 (73.96\%) & 848,490 \\
Post-Launch & 226,740 (27.61\%) & 594,251 (72.39\%) & 820,991 \\ \hline
Average & 200,370 (26.70$\pm$0.82\%) & 549,502 (73.30$\pm$0.82\%) & 749,872 \\
\hline
\end{tabular}
\caption{Distribution of account identities (bot vs human) across the three vaccine rollout periods. Percentages are calculated relative to the total number of users in each period.}
\label{tab:distribution_account_identities}
\end{table}

\begin{table}[h]
\centering
\begin{tabular}{p{3cm}p{2cm}p{2cm}p{2cm}p{2cm}}
\hline
~ & \textbf{All} & \textbf{Pre \newline Vaccine} & \textbf{Vaccine \newline Launch} & \textbf{Post \newline Launch} \\
\hline
Total Tweets & 8,676,375 & 2,257,045 & 3,425,624 & 2,993,706 \\
~ & ~ & ~ & ~ & ~ \\ 
Tweets with low-credibility narratives & 5,890,967 & 1,571,283 & 2,423,270 & 1,896,414 \\
(\%) & 67.89 & 69.62 & 70.74 & 63.35 \\

~ & ~ & ~ & ~ & ~ \\ 
Bot Tweets & 3,556,465 & 917,289 & 1,419,708 & 1,219,468 \\
Bot posted tweets with low-credibility narratives & 2,674,907 & 711,584 & 1,098,949 & 896,374 \\
(\%) & 75.21 & 77.57 & 77.41 & 73.51 \\

~ & ~ & ~ & ~ & ~ \\ 

Human Tweets & 5,119,910 & 1,339,756 & 2,005,916 & 1,654,238 \\
Human posted tweets with low-credibility narratives & 3,216,056 & 859,699 & 1,324,321 & 1,032,040 \\
(\%) & 64.25 & 64.17 & 66.02 & 58.16 \\
\hline
\end{tabular}
\caption{Distribution of misinformation tweets across the COVID-19 vaccine rollout periods, separated by bot and human accounts. Percentages represent the proportion of misinformation tweets relative to the total tweets produced by each account type.}
\label{tab:narratives_distribution}
\end{table}

\begin{table}[h]
\centering
\caption{Descriptive Statistics (Mean $\pm$ Standard Deviation)}
\begin{tabular}{lcc}
\hline
~ & \textbf{Appeal} & \textbf{Scope} \\
\hline
Low-credibility narratives (Bot)    & $0.12 \pm 6.33$   & $63.25 \pm 4048.22$ \\
Low-credibility narratives (Human)  & $2.81 \pm 95.5586$  & $5429.98 \pm 719620.90$ \\
Non-low-credibility narratives (Bot)       & $0.17 \pm 5.88$   & $47.29 \pm 777.416$ \\
Non-low-credibility narratives (Human)     & $76.43 \pm 2551.03$ & $164296.51 \pm 18444870$ \\
\hline
\end{tabular}
\label{tab:appeal_scope_desc}
\end{table}

\section{Narrative Groups}
\label{supp:narrative_groups}
\autoref{tab:narrative_groups} presents the proportion of low-credibility narratives groups in the CovidInfo dataset and illustrative messages.

\begin{table}[h]
    \centering
    \begin{tabular}{p{3cm}p{3cm}p{6cm}}
    \hline
       \textbf{Narrative Group} & \textbf{Percentage} & \textbf{Illustrative Messages} \\ \hline
       Fake cure & 97.13 & salt solution can cure covid19 \\ 
       Conspiracy & 2.68 & the subsequent vaccine as a genocidal weapon to kill us \\ 
       False fact or prevention & 0.21 & Russia says no to booze after vaccine shot!! \\ 
       Fake treatment & 0.001 & Immediately add immunity-building/anti-inflammatory/anti-viral garlic/Vitamin D3 to the treatment mix! Garlic cuts colds by 50\% (COVID-19 is a form of a cold) \\ 
       False public health responses & 0.0001 & The purpose of 'track \& trace' is for governments to use the manufactured 'Corona crisis' to install Orwellian Police State controls.  \\
    \hline
    \end{tabular}
    \caption{Proportion of narrative groups of the low-credibility narratives from the CovidInfo dataset and illustrative messages}
    \label{tab:narrative_groups}
\end{table}

\section{Matching Messages}
\label{supp:matching_messages}
\autoref{tab:matching_messages} presents the matching messages from using the cosine similarity message matching algorithm.

\begin{table}[h]
    \centering
    \begin{tabular}{p{6cm}p{6cm}}
        \hline 
        \textbf{Original message (from \cite{memon2020characterizing}}) & \textbf{Matching message (from COVID data)} \\ \hline 
         californian dies hours after getting covid-19 vaccine & covid19 vaccine is a conspiracy to kill people in order to get more money from patients \\ 
         this seems too perfect a bioweapon to have occurred naturally and the path from canadian lab to wuhan & wuhan coronavirus is an engineered offensive bioweapon. you're not being told this \\
         can sesame oil cure the coronavirus? how dangerous is it compared to the flu? & first article i see this morning, cocaine cures the coronavirus \\
         rinsing your nose with salt water will help contain the coronavirus & chewing garlic, mineral drinks, disinfectant, nasal sprays, these are some of the cures for the coronavirus \\
        \hline
     \end{tabular}
    \caption{Matching messages}
    \label{tab:matching_messages}
\end{table}

\section{Tweedie Regression Models}
\label{supp:tweedie_models}
\autoref{tab:tweedie_results} presents the full results for the Tweedie Regression Models for the Appeal and Scope metrics.

\begin{table}[h]
\centering
\resizebox{\textwidth}{!}{
\begin{tabular}{lcccc}
\hline
 & \multicolumn{2}{c}{\textbf{Baseline Model}} & \multicolumn{2}{c}{\textbf{Conditional Effect Model}} \\
\cline{2-5}
 & Appeal & Scope & Appeal & Scope \\
\hline

Null Deviance & 120,725,423 & $1.5921\times10^{11}$ & 120,725,423 & $1.592\times10^{11}$ \\
Residual Deviance & 58,602,020 & $6.03\times10^{10}$ & 58,564,257 & $6.023\times10^{10}$ \\
Degrees of Freedom & 13,938,867 & 13,938,876 & 13,938,876 & 13,938,876 \\
Dispersion Parameter & 398.65 & 7,616,587 & 386.09 & 7,580,770 \\
$R^2$ & 0.51 & 0.10 & 0.51 & 0.10 \\ 

\hline
\textbf{Coefficients: Standard Error (Effect Size)} & \multicolumn{4}{c}{} \\
\hline

Intercept & 0.0392$^{***}$ (1.16\%) & -- & 3.889$^{***}$ (1.11\%) & -- \\

IsBot=True & -2.42$^{***}$ (-91.11\%) & -2.26$^{***}$ (-89.56\%) & -2.19$^{***}$ (-88.81\%) & -1.64$^{***}$ (-80.60\%) \\

Narrative: Fake Cure & 0.074$^{***}$ (7.68\%) & 0.611$^{***}$ (84.23\%) & 0.074$^{***}$ (7.65\%) & 0.616$^{***}$ (85.15\%) \\

Narrative: Fake Treatment & -2.314 (-90.11\%) & -3.11 (-95.54\%) & -2.33 (-90.27\%) & -3.15 (-95.71\%) \\

Narrative: False Fact/Prevention & -0.333$^{***}$ (-28.34\%) & -0.0004 (-0.04\%) & -0.338$^{***}$ (-28.68\%) & -0.428$^{***}$ (-34.82\%) \\

Narrative: False Public Health Response & -2.281 (-89.78\%) & -2.37 (-90.65\%) & -2.37 (-90.65\%) & -2.40 (-90.93\%) \\

IsRetweet=True & -5.132$^{***}$ (-99.41\%) & -3.37$^{***}$ (-96.56\%) & -5.13$^{***}$ (-99.41\%) & -3.38$^{***}$ (-96.60\%) \\

Time Period = 2 & 0.067$^{***}$ (6.92\%) & 0.419$^{***}$ (52.04\%) & 0.132$^{***}$ (14.11\%) & 0.618$^{***}$ (85.52\%) \\

Time Period = 3 & 0.0438$^{**}$ (4.47\%) & 0.225$^{***}$ (25.23\%) & 0.0959$^{***}$ (10.06\%) & 0.380$^{***}$ (46.23\%) \\

Account Age & 2.31E-4$^{***}$ (0.023\%) & 4.79E-4$^{***}$ (0.048\%) & 2.31E-4$^{***}$ (0.023\%) & $4.83\times10^{-4}$$^{***}$ (0.048\%) \\

IsBot $\times$ TimePeriod = Vaccine Launch & -- & -- & -0.359$^{***}$ (-30.16\%) & -0.976$^{***}$ (-62.32\%) \\

IsBot $\times$ TimePeriod = Post Launch & -- & -- & -0.278$^{***}$ (-24.27\%) & -0.684$^{***}$ (-49.54\%) \\

\hline
\multicolumn{5}{l}{\footnotesize Notes: $^{***}$ p$<$0.001, $^{**}$ p$<$0.01, $^{*}$ p$<$0.05.} \\
\end{tabular}}
\caption{Tweedie Regression Results for Appeal and Scope}
\label{tab:tweedie_results}
\end{table}

\section{VIF Values}
\label{supp:vif_values}
We calculated the VIF values for the predictors in the Tweedie regression models. \autoref{tab:vif_baseline} presents the VIF values for the Baseline Model, and \autoref{tab:vif_conditional} presents the VIF values for the Conditional Effect Model. All adjusted VIF values are close to 1, indicating no evidence of problematic multicollinearity among predictors.

\begin{table}[h]
\centering
\begin{tabular}{lccc}
\hline
\textbf{Variable} & \textbf{GVIF} & \textbf{Df} & \textbf{GVIF$^{1/(2\times Df)}$} \\
\hline
is\_bot & 1.040030 & 1 & 1.019819 \\
narrative\_group & 1.001182 & 4 & 1.000148 \\
is\_retweet & 1.013359 & 1 & 1.006658 \\
time\_period & 1.001865 & 2 & 1.000466 \\
account\_age & 1.035558 & 1 & 1.017624 \\
\hline
\end{tabular}
\caption{Variance Inflation Factor (VIF) diagnostics for predictors in the Tweedie regression model for the \textbf{Baseline Model}. All adjusted VIF values (GVIF) are close to 1, indicating no evidence of problematic multicollinearity among predictors.}
\label{tab:vif_baseline}
\end{table}

\begin{table}[h]
\centering
\begin{tabular}{lccc}
\hline
\textbf{Variable} & \textbf{GVIF} & \textbf{Df} & \textbf{GVIF$^{1/(2\times Df)}$} \\
\hline
is\_bot & 3.486228 & 1 & 1.867144 \\
narrative\_group & 1.001159 & 4 & 1.000145 \\
is\_retweet & 1.013217 & 1 & 1.006587 \\
time\_period & 1.540943 & 2 & 1.114158 \\
account\_age & 1.036064 & 1 & 1.017872 \\
is\_bot:time\_period & 4.335661 & 2 & 1.442992 \\
\hline
\end{tabular}
\caption{Variance Inflation Factor (VIF) diagnostics for predictors in the Tweedie regression model for the \textbf{Conditional Effect Model}. Adjusted VIF values (VIF) remain below commonly accepted thresholds, indicating no problematic multicollinearity.}
\label{tab:vif_conditional}
\end{table}


\nolinenumbers

\end{document}